\documentclass[conference]{IEEEtran}
\usepackage[papersize={8.5in,11in},left=0.65in,right=0.65in,top=0.751in,bottom=1.2in]{geometry}
\IEEEoverridecommandlockouts
\usepackage{cite}
\usepackage{amsmath,amssymb,amsfonts}
\usepackage{algorithmic}
\usepackage{graphicx}
\usepackage{textcomp}
\usepackage{xcolor}
\def\BibTeX{{\rm B\kern-.05em{\sc i\kern-.025em b}\kern-.08em
    T\kern-.1667em\lower.7ex\hbox{E}\kern-.125emX}}
\graphicspath{{./Figures/}} 
 \usepackage{subcaption}
 \usepackage{url}
 \usepackage{mathtools} 

\newcommand{\bm}{\mathbf}
\newcommand{\be}{\begin{equation}}
\newcommand{\ee}{\end{equation}}
\newcommand{\bea}{\begin{eqnarray}}
\newcommand{\eea}{\end{eqnarray}}

\newcommand{\z}{{\bm z}}

\newcommand{\br}{{\bm r}}

\newcommand{\bA}{{\bm A}}
\newcommand{\bI}{{\bm I}}

\newcommand{\bF}{{\bf F}}
\newcommand{\bG}{{\bf G}}
\newcommand{\bD}{{\bf D}}

\newcommand{\bH}{{\bf H}}

\newcommand{\bZ}{{\bf Z}}

\newcommand{\bg}{{\bf g}}
\newcommand{\bu}{{\bf u}}

\newcommand{\bd}{{\bf d}}
\newcommand{\bs}{{\bf s}}

\begin{document}
\title{Time Frequency Localized Pulse for Delay Doppler Domain Data Transmission }
\author{\IEEEauthorblockN{Sanoopkumar P. S\IEEEauthorrefmark{1},
Muyiwa Balogun\IEEEauthorrefmark{2},
Liam Barry\IEEEauthorrefmark{2}, and 
Arman Farhang\IEEEauthorrefmark{1}}
\IEEEauthorblockA{\IEEEauthorrefmark{1}Department of Electronic \& Electrical Engineering, Trinity College Dublin, Ireland}

\IEEEauthorblockA{\IEEEauthorrefmark{2}School of Electronic Engineering, Dublin City University, Dublin, Ireland\\
Email: \{pungayis, arman.farhang\}@tcd.ie.,
\{balogun.muyiwablessing, liam.barry\}@dcu.ie}
\thanks{
This publication has emanated from research conducted with the financial support of Taighde Éireann – Research Ireland under Grant numbers 19/FFP/7005(T) and 21/US/3757. 
}
}
\maketitle
\begin{abstract}
Orthogonal time frequency space (OTFS) is a strong candidate waveform for sixth generation wireless communication networks (6G), which can effectively handle time varying wireless channels. In this paper, we analyze the effect of fractional delay in delay Doppler (DD) domain multiplexing techniques. We develop a vector-matrix input-output relationship for the DD domain data transmission system by incorporating the effective pulse shaping filter between the transmitter and receiver along with the channel. Using this input-output relationship, we analyze the effect of the pulse shaping filter on the channel estimation and BER performance in the presence of fractional delay and uncompensated fractional timing offset (TO). For the first time, we propose the use of time-frequency localized (TFL) pulse shaping for the OTFS waveform to overcome the interference due to fractional delays.   We show that our proposed TFL-OTFS outperforms the widely used raised cosine pulse-shaped OTFS (RC-OTFS) in the presence of fractional delays. Additionally, TFL-OTFS also shows very high robustness against uncompensated fractional TO, compared to RC-OTFS. 

\end{abstract}
\section{Introduction}
Recent advances in autonomous driving, high-speed trains and convergence of satellite and terrestrial networks call for a highly reliable and robust air interface that can cope with the underlying time-varying channels\cite{Liu_2024}.
The main challenge in achieving highly reliable and low latency communication in high mobility scenarios is the time and frequency selectivity of the wireless channel \cite{v2e}. Orthogonal frequency division multiplexing (OFDM) is used as the air interface in most of the current wireless standards \cite{3gppr16}. Although OFDM is inherently robust to frequency selectivity, the time selectivity of the channel destroys the orthogonality between sub-carriers up to a level that breaks the communication link \cite{tse}. To tackle this issue, orthogonal time frequency space (OTFS) modulation scheme presented the novel idea of data transmission in the delay-Doppler (DD) domain where the doubly-selective channel has a 2D time-invariant and sparse representation \cite{hadani_2017}. The DD domain data transmission schemes outperform conventional time-frequency (TF) domain data transmission by harvesting the full diversity of the wireless channels\cite{raviteja_MP}.  

In DD domain, the wireless channel can be efficiently estimated using a single impulse pilot embedded with data and sufficient guards across delay and Doppler dimensions \cite{raviteja_embedded}. When the delays and Doppler shifts of the channel are integer multiple of delay Doppler spacing of OTFS, the guards required for impulse pilot would be limited by the delay and Doppler spread of the channel\cite{raviteja_embedded}. However, in realistic scenarios, the channel delay and Doppler shifts are not integer multiples of delay and Doppler spacing of OTFS, thereby resulting in fractional delay and Doppler \cite{arslan_2022,raviteja_embedded}. In such scenarios, the impulse pilot-based channel estimation would require an entire OTFS frame for channel estimation, which is not a viable solution. 

In the presence of fractional delay the transmitted signal will spread along the delay dimension beyond the delay spread of the channel \cite{zak_1}. Similarly, the fractional Doppler will spread the transmitted signal along the Doppler dimension beyond the Doppler spread of the channel \cite{raviteja_MP}. Hence in the presence of both fractional delay and Doppler, the transmitted symbols in each DD grid will interfere with the symbols in the entire DD grid. For the embedded pilot-based channel estimation techniques, full guards across the Doppler dimension is used in fractional Doppler scenario \cite{raviteja_embedded}. However, if we use full guard across the delay dimension to counteract the fractional delay, the spectral efficiency of the system will reduce drastically as the number of delay bins is usually much larger than Doppler bins. To reduce the effect of fractional delays, a suitable transmit pulse which is very localized across delay dimensions is required \cite{arslan_2022}. Pulse shaping for DD domain data transmission is addressed in \cite{oddm_2022,saifkhan_2023_part2,Mohsen_gen}. In \cite{oddm_2022} a train of truncated raised cosine (RC) pulse is used to transmit information symbols in DD domain. In \cite{saifkhan_2023_part2}, authors used a two-dimensional transmit pulse to develop a single-stage OTFS transceiver. They presented the results for both RC and sinc pulse across both delay and Doppler dimensions. A generalized framework for pulse shaping in DD domain is presented in \cite{Mohsen_gen}. All of the aforementioned works use either RC pulse or sinc pulse for pulse shaping. Additionally, they also did not investigate the effect of fractional delay on channel estimation. Channel estimation for OTFS systems in the presence of fractional delays is considered in \cite{Saif_Khan_frac_2023,Sai_2023,Yogesh_2024,Rao_frac}. While an embedded impulse pilot is used in \cite{Saif_Khan_frac_2023,Sai_2023,Yogesh_2024}, an intermittent pilot frame which occupies the entire OTFS DD plane is used in \cite{Rao_frac}. However, due to fractional delay, iterative interference cancellation has to be applied with the channel estimation procedure in methods presented in \cite{Saif_Khan_frac_2023,Sai_2023,Yogesh_2024,Rao_frac}. 

One of the main advantages of OTFS is the ease of channel estimation. However in a practical wireless system with fractional delays, one can not enjoy this advantage offered by the OTFS using existing transceiver techniques.  Existing DD domain data transmission using OTFS in \cite{hadani_2017,oddm_2022,saifkhan_2023_part2,Mohsen_gen} does not address the effect of fractional delays. On the other hand, the existing works in literature which address fractional delays require computationally complex iterative techniques to deal with the interference resulting from fractional delays \cite{Saif_Khan_frac_2023,Sai_2023,Yogesh_2024,Rao_frac}. This motivated us to address the problem at its roots, i.e., the pulse shape, and to develop an OTFS transmission that reduces the spread due to fractional delay.

In \cite{Haas1997time}, authors proposed a pulse that is well localized in TF domain and creates less interference than the RC for OFDM systems. In \cite{Farhang_underwater}, this pulse was used for data transmission in underwater acoustic channels. In this paper, we propose the use of this time-frequency localized (TFL) pulse for data transmission in DD domain. To the best of our knowledge, this is the first attempt in the literature to reduce the effect of fractional delay in OTFS systems. For the sake of simplicity in explanation, we call the proposed method TFL-OTFS whereas the existing RC-based OTFS transmission is referred to as RC-OTFS.  
We develop a novel matrix vector input-output  relationship for DD domain data transmission by incorporating transmit and receive pulse shaping filters. Using this novel input-output relation we analyze the DD domain interference for both RC-OTFS and TFL-OTFS and show that TFL-OTFS has much lesser interference in the DD domain than RC-OTFS. We also develop a DD domain transceiver for TFL-OTFS using a discrete implementation of the TFL pulse. We investigate the performance of embedded pilot-based channel estimation and show that the proposed TFL-OTFS outperforms RC-OTFS in the presence of both fractional delay and Doppler. Furthermore, the robustness of the proposed TFL against uncompensated fractional TO is also investigated. We show that TFL-OTFS outperforms the RC-OTFS even when the system has residual uncompensated TO.


The rest of the paper is organized as follows. Section \ref{sec:system}, presents the system model with the novel vector matrix input-output relationship. The effect of fractional delay in RC-based OTFS is described in Section \ref{sec:2}. Section \ref{sec:tfl}, presents the proposed TFL-based OTFS. The simulation results are shown in Section \ref{sec:sim}. Finally, the paper is concluded in Section \ref{sec:conl}. 

\textit{Notations:} In this paper, normal letters, boldface lowercase, and boldface uppercase letters are used to represent scalar values, vectors, and matrices, respectively. 
The Hadamard product and Kronecker product are represented by $\odot$ and $\otimes$, respectively. Circular convolution is denoted by $\circledast$. The function ${\rm vec}\left({\bA}\right)$ forms a vector by concatenating the columns of the matrix ${\bf A}$. Operator $(.)_i$ denotes modulo with respect to $i$.The Dirac delta function is represented as where $\delta(.)$. The normalized $N$-point discrete Fourier transform (DFT) matrix is represented as $\bF_N$ with the elements ${F}_N[m,n]= 1/{\sqrt{N}} e^{-j2 \pi \frac{mn}{N}}$ for $m,n=0,\ldots, N-1$.
\section{System Model}\label{sec:system}
In this paper, we consider an OTFS system with $M$ delay bins and $N$ Doppler bins. With a sampling rate of $T_{\rm s}$ seconds the system has a delay resolution of $T_{\rm s}$ seconds and Doppler resolution of $\frac{1}{MNT_{\rm s}}$ Hz. 
At the OTFS transmitter, the information bits are first mapped onto QAM symbol constellation. The modulated symbols are stacked into an $M\times N$ matrix, $\bD$, whose elements $D[m,n]$ represent the data symbols on the delay bin $m$ and the Doppler bin $n$ on the grid. The DD domain symbols are converted to delay time (DT) domain using inverse discrete Zak transform \cite{zak_1}
\be \label{eqn:zak1}
\bs={\rm IDZT}(\bD)={\rm vec}(\bD \bF_N^{\rm H})=(\bF_N^{\rm H}\otimes \bI_M) \bd,
\ee
where $\bd=\rm{vec}(\bD)$. The last $M_{\rm cp}$ samples of $\bs$ are then appended at the beginning of $\bd$ as a cyclic prefix (CP) to form the OTFS signal  $\bs_{\rm cp}$.
After adding CP, the discrete time OTFS signal $\bs_{\rm{cp}}$ is pulse-shaped with the transmit pulse $p_{\rm tx}(t)$ as,
\be\label{eqn:tx_cont}
s_{\rm cp}(t)=\sum_{\kappa=0}^{M_{\rm cp}+NM-1}s_{\rm cp}[\kappa]p_{\rm tx}(t-\kappa T_{\rm s}).
\ee
The continuous-time signal $s_{\rm cp}(t)$ is then up-converted to the carrier frequency $f_{\rm c}$ and transmitted over the wireless channel.  At the receiver, the received signal is down-converted to baseband which can be represented as
\be\label{eqn:rx_cont}
r(t)=\int_{-\infty}^{\infty}\int_{-\infty}^{\infty}h(\tau,\nu)s_{\rm cp}(t-\tau)e^{j2\pi\nu t}d\tau d\nu + \eta(t), 
\ee
where $h(\tau,\nu)$ is the DD domain channel response and $\eta(t)$ is the additive white Gaussian noise with the variance $\sigma^2$. 
Due to the sparse nature of the channel in DD domain, it can be expressed as  
\be \label{eqn:dd_channel}
h(\tau,\nu)=\sum_{i=1}^P h_i\delta(\tau-\tau_i)\delta(\nu-\nu_i),
\ee
where $h_i$, $\tau_i$ and $\nu_i$ are the complex channel gain, delay and Dopler shift of the $i^{\text{th}}$ path respectively.
 Using (\ref{eqn:dd_channel}) and (\ref{eqn:tx_cont}), $r(t)$ can be expressed as 
\be
r(t)=\sum_{i=1}^{P}h_i\!\!\!\!\!\!\!\sum_{\kappa=0}^{M_{\rm cp}+NM-1}\!\!\!\!\!\!\!\!s_{\rm cp}[\kappa]p_{\rm tx}(t-\tau_i-\kappa T_{\rm s})e^{j2\pi\nu_it}+\eta(t).
\ee
At the receiver, $r(t)$ is passed through the receive filter $p_{\rm rx}(t)$, which is matched to the transmit filter $p_{\rm tx}(t)$, i.e., $p_{\rm rx}(t)=p^*_{\rm tx}(-t)$. The output of the matched filter (MF) can be expressed as
\begin{align}\nonumber
   & y_{\rm MF}(t)=\sum_{i=1}^{P}h_i\!\!\!\!\!\!\!\sum_{\kappa=0}^{M_{\rm cp}+NM-1}\!\!\!\!\!\!\!\!s_{\rm cp}[\kappa]\\
    &\int_{-\infty}^{\infty}p_{\rm tx}(\tau-\tau_i-\kappa T_{\rm s})e^{j2\pi\nu_i\tau}p^*_{\rm tx}(\tau-t)d\tau+\eta(t).
\end{align}

By changing the variables in the integral and using the assumption that the maximum Doppler shift of the channel is smaller than the bandwidth of the filters (this is valid as a large number of subcarriers are used in practical systems \cite{zak_1}), the output of the MF can be expressed as, 


\begin{align}\nonumber
   y_{\rm MF}(t)\approx&\sum_{i=1}^{P}h_i\!\!\!\!\!\!\!\sum_{\kappa=0}^{M_{\rm cp}+NM-1}\!\!\!\!\!\!\!\!\bs_{\rm cp}[\kappa]
    e^{j2\pi\nu_i(\tau_i+\kappa T_{\rm s})}g(t-\kappa T_{\rm s}-\tau_i)\\&+\eta(t),
\end{align}
where $g(t)=\int_{-\infty}^{\infty}p_{\rm tx}(\tau)p^*_{\rm rx}(\tau-t)d\tau$ is the effective pulse between the transmitter and the receiver. Finally, the discrete-time received signal samples are obtained by sampling $y_{\rm MF}(t)$ at $T_{\rm s}$ intervals. Thus, the $\ell^{\text{th}}$ sample of the received signal can be expressed as 
\be \label{eqn:dt_sample}
y[\ell ]=\sum_{i=1}^{P}h_i'\!\!\!\!\!\!\!\sum_{\kappa=0}^{M_{\rm cp}+NM-1}\!\!\!\!\!\!\!\!s_{\rm cp}[\kappa]
    e^{j2\pi\frac{\epsilon_i\ell}{MN}}g_i[\ell-\kappa]+\eta[\ell],
\ee
where, $0\leq \ell\leq MN+M_{\rm cp}-1$, $h_i'=h_ie^{j2\pi\nu_i\tau_i}$, $\epsilon_i=\frac{\nu_i}{MNT_{\rm s}}$ and $g_i[\kappa]=g(\kappa T_{\rm s}-\tau_i)$ .
\subsection{Effective Channel in DD domain}
After discarding CP, the linear convolution in (\ref{eqn:dt_sample}) becomes circular convolution, i.e.,
\be\label{eqn:r_circ_conv}
r[\ell]=\sum_{i=1}^{P}h_i'\!\!\!\!\sum_{\kappa=0}^{NM-1}\!\!\!\!s[\kappa]
    e^{j2\pi\frac{\epsilon_i\ell}{MN}}g_i[(\ell-\kappa)_{MN}]+\eta[\ell]
\ee
where, $M_{\rm cp}\leq \ell\leq MN+M_{\rm cp}-1$. Stacking the signal samples $z[m]$ in a vector, (\ref{eqn:r_circ_conv}) can be represented in a more compact form as 
\be
\br=\sum_{i=1}^{P}h_i'(\bs\odot\bu_i)\circledast \bg_i,
\ee
where $u_i[n]=e^{j2\pi\frac{\epsilon_in}{MN}}$.
After performing the discrete Zak transform (DZT) (limiting to the fundamental period \cite{zak_1}) on $\br$, the received DD domain signal in the $m^{\text{th}}$ delay and $n^{\text{th}}$ Doppler bin, $Z[m,n]$ is  
\begin{align}\nonumber
Z[m,n]=\sum_{i=1}^{P}h_i'\sum_{l=0}^{M-1}\sum_{k=0}^{N-1}D[l,k]U_{{\rm z},i}[l,n-k]\\G_{{\rm z},i}[m-l,n]e^{-j2\pi\frac{n}{N}\rho_{m-l}}\label{eqn:zak_rxr}
\end{align}
where, $rho_p=0$ for $p\geq 0$ and $rho_p=1$ for $p < 0$.
In (\ref{eqn:zak_rxr}), $U_{{\rm z},i}[m,n]$ and $G_{{\rm z},i}[m,n]$ are the elements of $M\times N$ matrices $\mathbf{U}_{{\rm z},i}$ and $\mathbf{G}_{{\rm z},i}$ respectively where $\mathbf{U}_{{\rm z},i}$ and $\mathbf{G}_{{\rm z},i}$ are the DZT of the sequences ${\bf u}_i$ and ${\bf g}_i$ respectively. 
For developing signal processing solutions we use the vector-matrix form of the input-output relation in DD domain. The signal samples in (\ref{eqn:zak_rxr}) can be stacked to obtain the DD domain received signal matrix, $\bf Z$. 
In matrix-vector notation $\z=\rm{vec}(\bZ)$ can be expressed as
\begin{align} 
    \z=\left(\sum_{i=1}^{P}h_i'\widetilde{\bm U}_i\odot\widetilde{\bG}_i\right)\bd=\bH_{\rm dd}^{\rm eff}\bd\label{eqn:dd_H_eff2}
\end{align}
where, $\widetilde{\bm U}_i^0=[({\bm U}_i^0)^{\rm T} ({\bm U}_i^1)^{\rm T}, \hdots, ({\bm U}_i^{N-1})^{\rm T}]^{\rm T}$ is the first column block of $MN\times MN $ block circulant matrix $\widetilde{\bm U}_i$ and ${\bm U}_i^p={\boldsymbol 1}_{N\times 1}\otimes ({\bm u}_{{\rm z},i}^p)^{\rm T}$.
In (\ref{eqn:dd_H_eff2}), the $MN\times MN$ matrix $\widetilde{\bG}_i$ is defined as $\widetilde{\bG}_i={\boldsymbol 1}_{1\times N}\otimes {{\bG}}_i$, where ${{\bG}}_i=[({\bm G}_i^0)^{\rm T} ({\bm G}_i^1)^{\rm T}, \hdots, ({\bm G}_i^{N-1})^{\rm T}]^{\rm T}$, with ${\bm G}_i^p={\underline{\bm {G}}}_i^p\odot {\bm E}_p$,  ${\underline{\bm {G}}}_i^p$ is the $M \times M$ circulant matrix with first column as ${\bm g}_{{\rm z},i}^p$ and 
\be
{\bm E}_p=\left[ \begin{tabular}{ccccc}
     $1$& $e^{\frac{-j2\pi p}{N}}$ & $e^{\frac{-j2\pi p}{N}}$ & $\hdots$ & $e^{\frac{-j2\pi p}{N}}$ \\
     $1$ & $1$ & $e^{\frac{-j2\pi p}{N}}$ & $\hdots$ & $e^{\frac{-j2\pi p}{N}}$\\
     $1$ & $1$ & 1 & $\hdots$ & $e^{\frac{-j2\pi p}{N}}$\\
     $\vdots$ & $\vdots$ & $\vdots$ & $\ddots$ & $\vdots$\\
     $1$ & $1$ & 1 & $1$ & $1$.\\
\end{tabular}\right]
\ee
\section{Effect of Fractional Delay}\label{sec:2}
The predominant part of the existing literature on signal processing for OTFS considers that the delays of the resolvable paths are exactly integer multiples of the sampling period. However, in realistic scenarios, the interference caused by by fractional delays is unavoidable. In (\ref{eqn:dt_sample}), $g_i[n]=g[nT_{\rm s}-\tau_i]$ is the sampled effective pulse, shifted in time by the delay of $i^{\text{th}}$ delay tap. When the delays are integer multiples of sampling time $T_{\rm s}$,  $g_i[n]$ will be samples of $g(t)$ at integer multiples of $T_{\rm s}$.  In practical systems, all channel paths may not be resolved which results in fractional delays, i.e.,  $g_i[n]$ need not be samples of $g(t)$ at integer multiples of $T_{\rm s}$.   Channel estimation and data detection for OTFS systems in the presence of fractional delays are considered in \cite{Saif_Khan_frac_2023,Sai_2023,Yogesh_2024,Rao_frac}.  The authors used square root raised cosine (SRRC) pulses in these works as transmit and receive filters. The SRRC transmitter filter is given by \cite{barry_2004}
\begin{align}\label{eqn:rc_tx}
     p_{\rm tx}(t)&=\frac{4\beta}{\pi \sqrt{T_{\rm s}}}\frac{f_\beta(t)}{1-(4\beta t/T_{\rm s})^2},
     \end{align}
 where,    
     \begin{align}
     f_\beta(t)&=\cos\left(\frac{(1+\beta)\pi t}{T_{\rm s}}\right)+\frac{T_{\rm s}}{4\beta t}\sin \left(\frac{(1-\beta)\pi t}{T_{\rm s} }\right),
\end{align}
and $\beta $ is the roll-off factor.
Using the matched filter at the receiver, i.e., $p_{\rm rx}(t)=p_{\rm tx}(-t)$, the effective pulse between the transmitter and receiver will be a RC pulse given by,
\begin{align}
    g(t)=\frac{\sin (\pi t/T_{\rm s})}{\pi t/T_{\rm s}}\frac{\cos (\beta \pi t /T_{\rm s})}{1-(2\beta t/T_{\rm s})^2}
\end{align}
\begin{figure}
    \centering
    \includegraphics[width=0.7\linewidth]{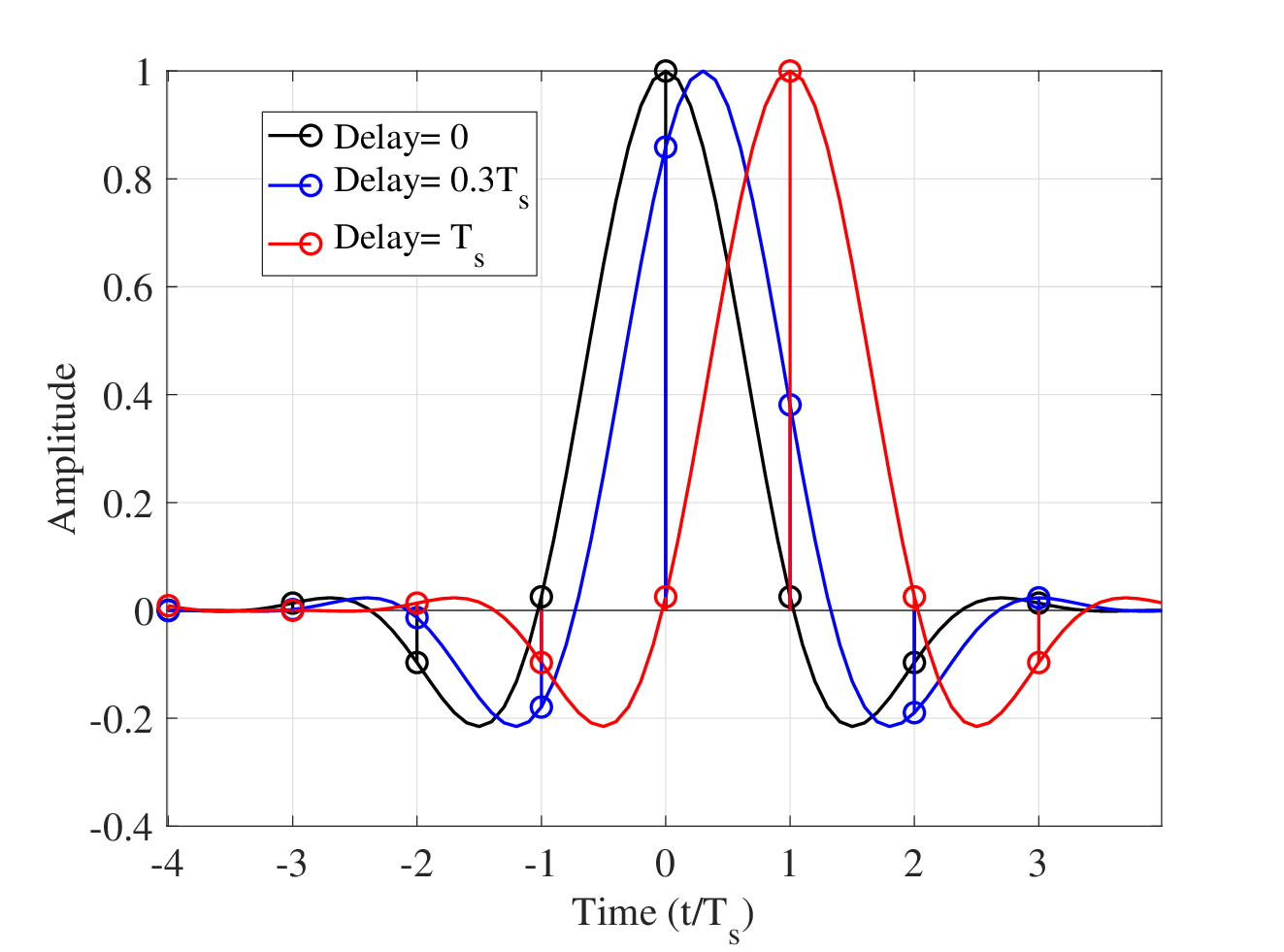}
    \caption{RC-OTFS ($\beta=0.22$) received pulses when an impulse is transmitted through a wireless channel with three paths having unit gain and zero Doppler shifts.}
    \label{fig:rc_frac1}
\end{figure}
In order to investigate the effect of integer and fractional delay, we consider a channel with three paths of $0$, $0.3 T_{\rm s}$ and $T_{\rm s}$ delays with unit gain and zero Doppler shift for each path. The roll-off factor for RC pulse in 5GNR is stipulated at $0.22$ \cite{5g_NR}, hence we consider an RC-OTFS with  $\beta= 0.22$. Fig. \ref{fig:rc_frac1} shows the received pulses in the time domain. The black, blue and red colors show the received pulse through paths with $0$, $0.3 T_{\rm s}$ and $T_{\rm s}$ delays, respectively. 
It can be observed that when there is an integer delay, the interference at sampling points is almost zero except in the peak sample. However, when there is a fractional delay, the peak does not occur at the sampling points and it creates significant interference at all other sampling points. The RC pulse can be made more localized in time by increasing the $\beta$, which can reduce the interference at sampling points for fractional delay scenarios. However, when the roll-off factor is increased, the signal spreads more in the frequency domain and degrades the system performance in the presence of fractional Doppler \cite{Mohsen_gen}.  
Due to the presence of fractional delay, computationally complex channel estimation techniques are required  to address the interference resulting from fractional delays \cite{Saif_Khan_frac_2023,Sai_2023,Yogesh_2024,Rao_frac}. To address this issue and to facilitate the use of low-complexity channel estimation methods, we propose the use of a TFL pulse for DD domain data transmission.  
\vspace{0 cm}
\section{Time frequency localized pulse shaping filter}\label{sec:tfl}
\vspace{0 cm}
\begin{figure}
    \centering
    \includegraphics[width=0.7\linewidth]{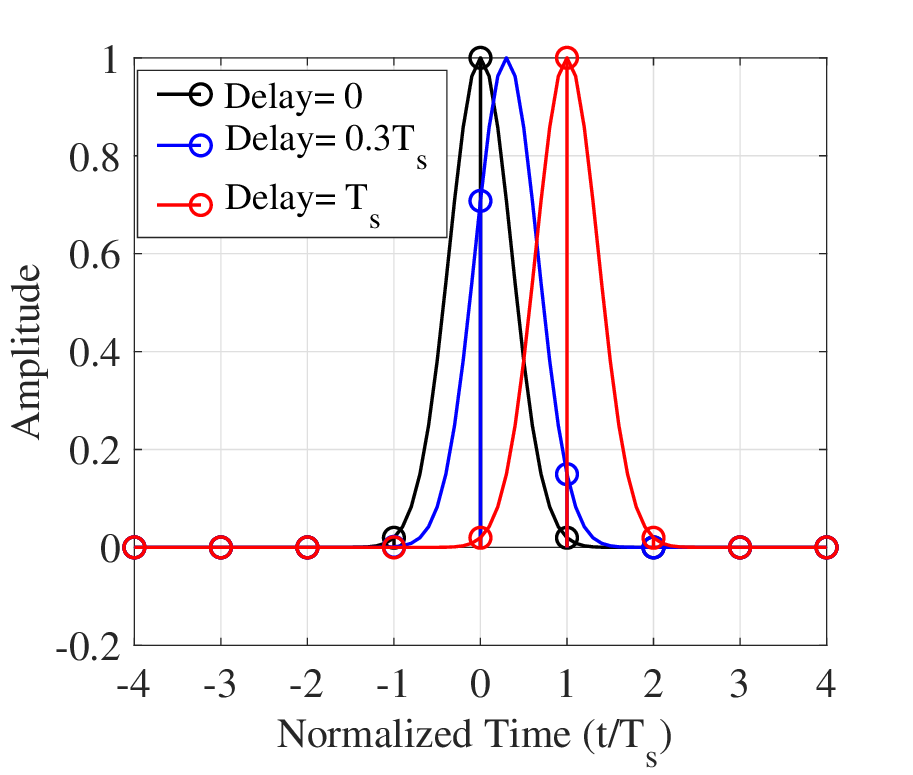}
    \caption{TFL-OTFS received pulses when an impulse is transmitted through a wireless channel with three paths having unit gain and zero Doppler shifts}
    \label{fig:gh_frac}
\end{figure}

Gaussian Hermite functions are widely used in engineering applications like image processing, optical communication, biomedical signal processing. In \cite{Haas1997time}, the authors proposed a TFL pulse for multicarrier communications using a linear combination of Gaussian Hermite functions, which can be described mathematically as
\begin{align}
     p_{\rm tx} (t)&=\sum_{p=0}^{4}D_{4p}\psi_{4p}(t),\label{eqn:her_fun}
\end{align}
where,
\begin{align}
    \psi_p(t)&=\frac{\mathcal{H}_p(t)}{\sqrt{2^p\,p!\,\sqrt{\pi}}}\,e^{-t^2/2},
\end{align}
and
\begin{align}
    \mathcal{H}_p(t)&=(-1)^pe^{t^2}\frac{{\rm d}^p}{{\rm d}t^p}e^{-t^2}. 
\end{align}

Using this TFL pulse we propose the TFL-OTFS which transmits the data in DD domain using the TFL pulse. Since the transmit pulse is real, the effective pulse after MF is given by $g_{\rm H}(t)=\int p_{\rm tx}(\tau)p_{\rm tx}(\tau-t)d\tau$. For the implementation of this proposed TFL-OTFS we make use of a discrete implementation of Gaussian Hermite pulse \cite{tri1, tri2}. 
The discrete  Gaussian Hermite pulses can be obtained as the eigenvectors of a symmetric Tridiagonal matrix $T_{\rm h}$ with the main diagonal 
\begin{equation}
    T_0[k']=-2\cos\!\!\left(\frac{\pi}{\sigma^2}\right)\!\sin\!\!\left(\frac{\pi k'}{\tilde{M}\sigma^2}\right)\!\sin\!\!\left(\frac{\pi}{\tilde{M}\sigma^2}((\tilde{M}-1)-k')\right),
\end{equation}
where $\tilde{M}$ is the number of sample points in the pulse, $0\leq k'\leq \tilde{M}-1$  and off diagonal elements as 
\begin{equation}
    T_1[k'']=\sin\left(\frac{\pi k''}{\tilde{M}\sigma^2}\right)\sin\left(\frac{\pi}{\tilde{M}\sigma^2}(\tilde{M}-k'')\right),
\end{equation}
for $1\leq k''\leq \tilde{M}-1$, i.e., 

\begin{align}
    T_{\rm h}=\left[ \begin{tabular}{cccccc}
        $T_0[0]$& $T_1[1]$& 0 & 0& $\hdots$ & 0\\
        $T_1[1]$& $T_0[1]$& $T_1[2]$ & 0& $\hdots$ & 0\\
        0& $T_1[2]$& $T_0[2]$ & $T_1[3]$& $\hdots$ & 0\\
        $\vdots$ &$\vdots$ &$\vdots$ &$\vdots$ &$\ddots$ & $\vdots$ \\
        0 & 0&  0 & 0& $\hdots$& $T_0[N-1]$
    \end{tabular}
    \right].
\end{align}
Eigen value decomposition of $\bf{T}_{\rm h}$  can obtain discrete Hermite functions, then (\ref{eqn:her_fun}) can be used to generate the transmit pulse. 

\begin{figure}
\begin{subfigure}{0.33\columnwidth}
    \includegraphics[width=1\columnwidth]{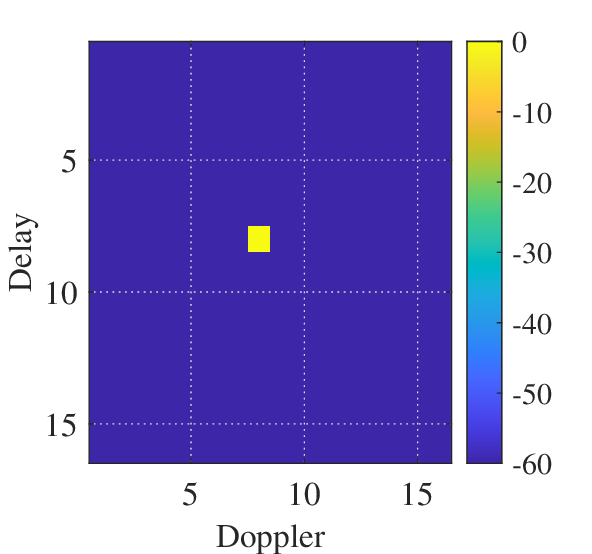}
    \caption{Transmitted Signal}
    \label{fig:dd1}
\end{subfigure}
\hspace{-0.3 cm}
\begin{subfigure}{0.33\columnwidth}
    \includegraphics[width=1\columnwidth]{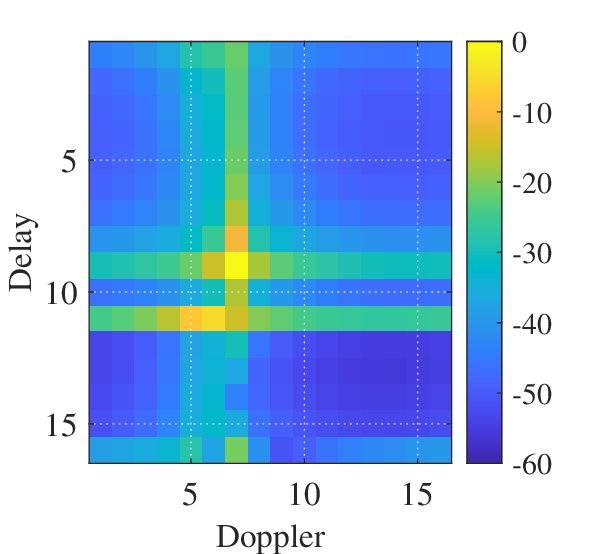}
    \caption{$\beta=0$}
    \label{fig:dd2}
\end{subfigure}
\hspace{-0.3 cm}
\begin{subfigure}{0.33\columnwidth}
    \includegraphics[width=1\columnwidth]{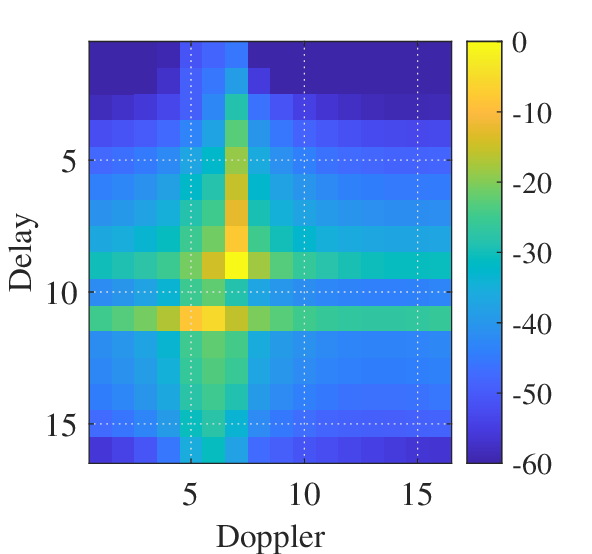}
    \caption{$\beta=0.2$}
    \label{fig:dd3}
\end{subfigure}
\vspace{-0.3 cm}
\begin{subfigure}{0.33\columnwidth}
    \includegraphics[width=1\columnwidth]{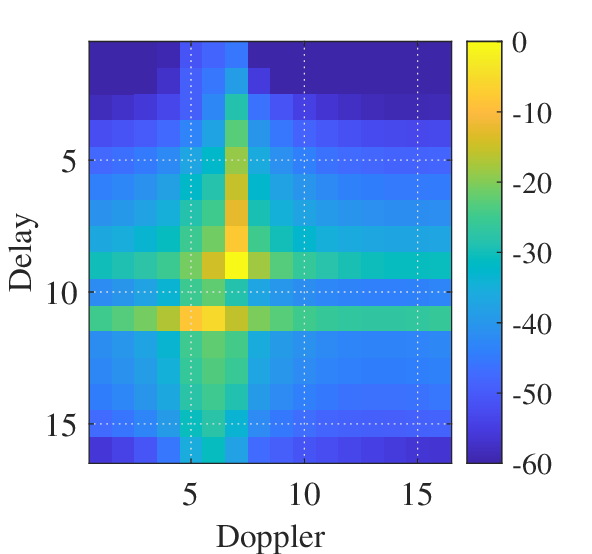}
    \caption{$\beta=0.5$}
    \label{fig:dd4}
\end{subfigure}
\hspace{-0.3 cm}
\begin{subfigure}{0.33\columnwidth}
    \includegraphics[width=1\columnwidth]{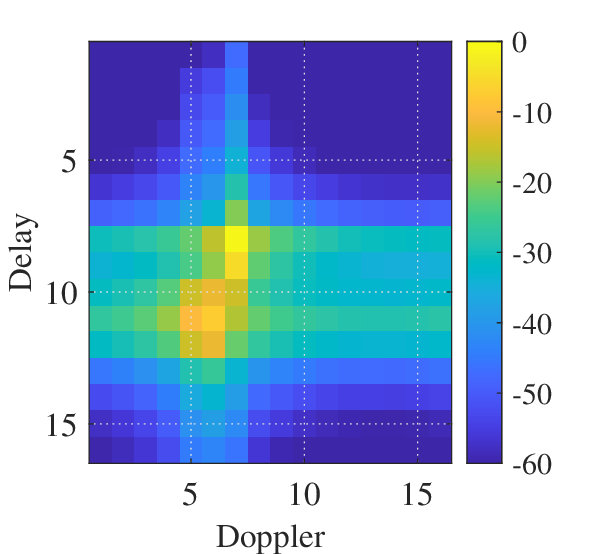}
    \caption{$\beta=1$}
    \label{fig:dd6}
\end{subfigure}
\hspace{-0.3 cm}
\begin{subfigure}{0.33\columnwidth}
    \includegraphics[width=1\columnwidth]{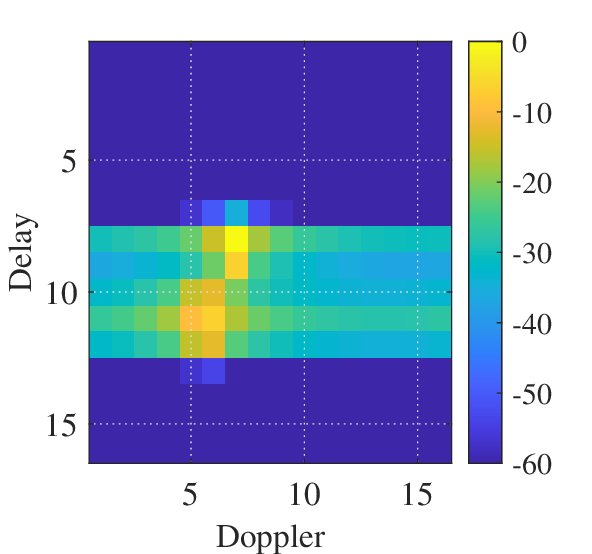}
    \caption{TFL}
    \label{fig:dd7}
 \end{subfigure}
    \vspace{0.3cm}\caption{Spreading of the (a) transmitted impulse in DD domain for (b-e) RC-OTFS and (f) TFL-OTFS.}
    \label{fig:dd_spread}
\end{figure}

The effect of fractional delay and integer delay for TFL pulse is shown in Fig.~\ref{fig:gh_frac} for the same wireless scenarios considered for RC-OTFS in Fig. \ref{fig:rc_frac1}. It can be seen that compared to RC-OTFS, TFL-OTFS is more localized in time resulting in smaller interference at sampling points with fractional delays. The interference due to fractional delay, in DD domain for RC-OTFS and TFL-OTFS are compared in Fig.~\ref{fig:dd_spread}. For a transmitted impulse signal in DD domain as shown in yellow color in Fig.~\ref{fig:dd1}, the received signal with RC pulse shaping for different values of roll-off factor is shown in Fig.~\ref{fig:dd2} and \ref{fig:dd6}. Finally, Fig.~\ref{fig:dd7} shows the received DD domain response with TFL-OTFS.  It can be seen that the impulse spreads throughout the DD grid for RC-OTFS whereas for TFL it is limited. Furthermore, it can be also seen that for RC-OTFS with smaller roll-off factors, the spread is very high compared to larger roll-off factors. 

\section{Simulation Results}\label{sec:sim}
In this section, we numerically analyze the performance of OTFS transmission with the TFL pulse and compare it with the RC pulse. We considered a system with $f_{\rm c}=5.9$ GHz, subcarrier spacing of $15$ kHz and bandwidth of $0.4$ MHz. The DD grid is divided into $M=32$, delay bins and $N= 16$, Doppler bins. The delay and Doppler resolution of the system are $2.1$~$\mu s$ and $937.5$ Hz respectively. 4-QAM constellation is used for data transmission. The wireless channel is simulated using $6$ paths with exponential power delay profile. The fractional delay for each tap normalized to delay resolution is randomly generated in each realization and is chosen from a uniform distribution with the range $[-0.5, 0.5]$. The pilot with cyclic prefix (PCP) is used for channel estimation as proposed in \cite{Sanoop_pcp}. PCP of length $9$  and pilot power of $30$~dB is embedded with the data frame in DD domain. Basis expansion model based interpolation is also used as described in \cite{Sanoop_pcp}. Unless specified otherwise we consider a mobile speed of $500$~km/h.

\begin{figure}
    \centering
    \includegraphics[width=0.75\linewidth]{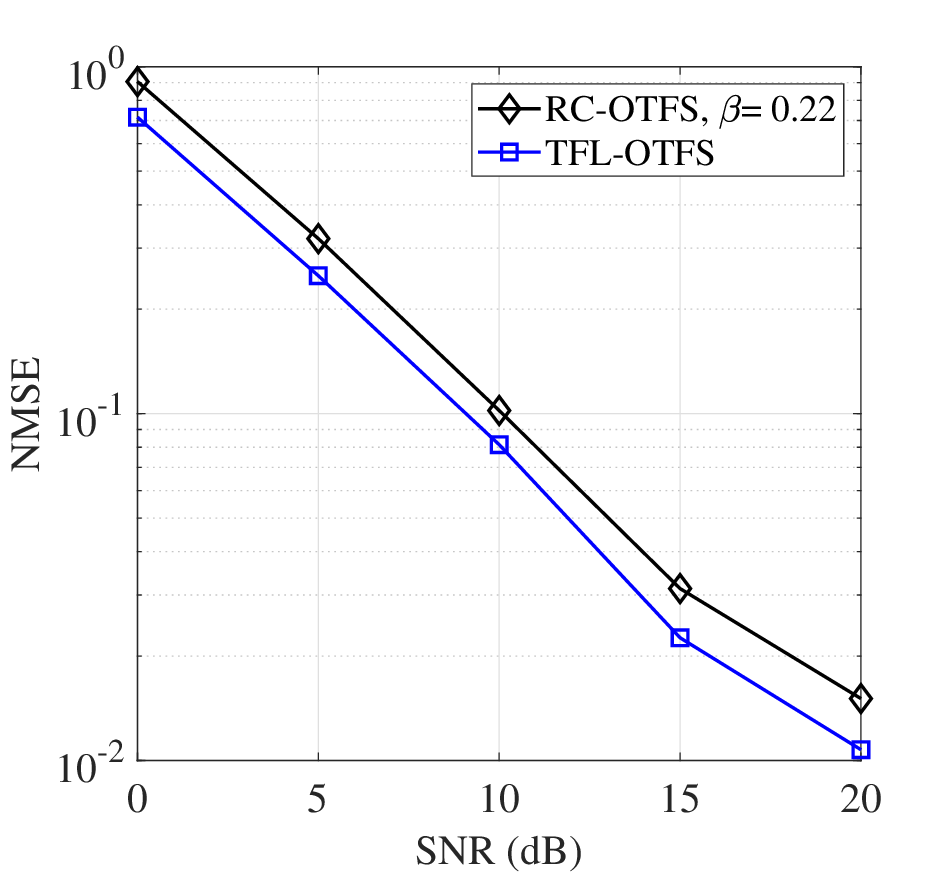}
    \caption{Channel estimation performance versus SNR.}
    \label{fig:nmse}
\end{figure}

\begin{figure}
    \centering
    \includegraphics[width=0.75\linewidth]{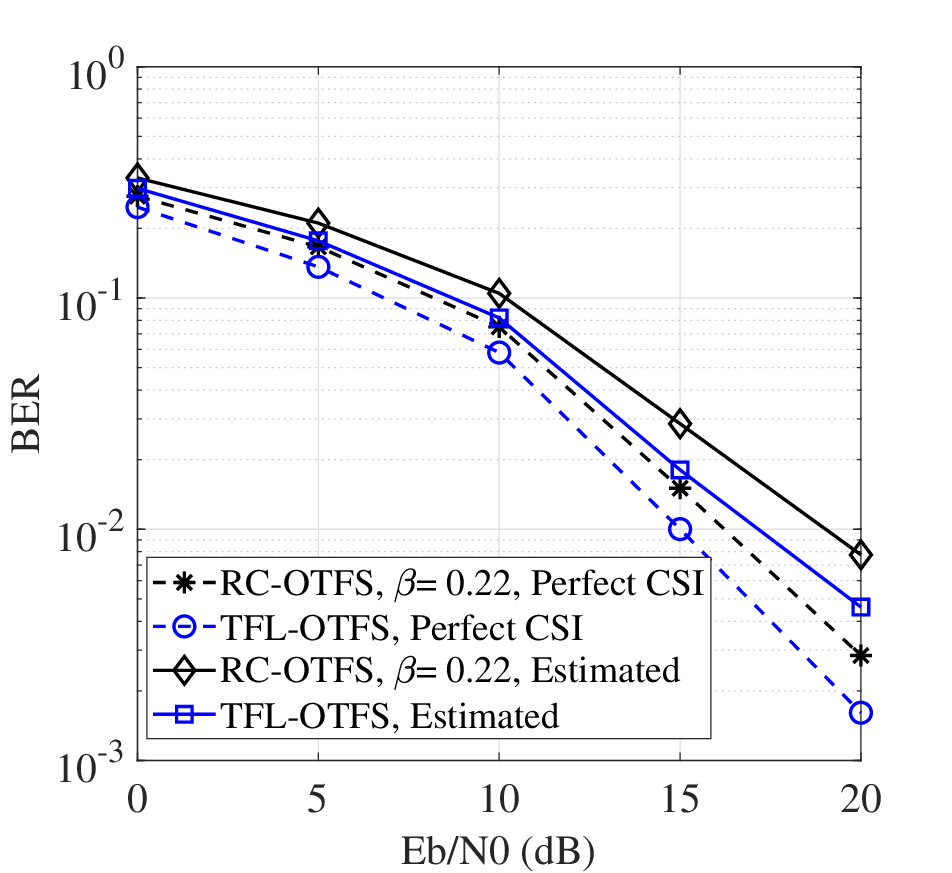}
    \caption{BER performance of MMSE detection.}
    \label{fig:ber}
\end{figure}

The RC pulse with $\beta=0.22$ is stipulated in 5G NR \cite{3gppr16}. Hence, we compare the performance of TFL-OTFS with RC-OTFS with $\beta=0.22$. In Fig. \ref{fig:nmse}, the normalized mean squared error (NMSE) of channel estimation is shown. The bit error rate performance of the minimum mean squared error (MMSE) detector for TFL- and RC-OTFS with perfect knowledge of the channel state information (CSI) and estimated CSI is shown in Fig. \ref{fig:ber}. As discussed in the previous section, RC-OTFS with small values of roll-off factor results in high interference in the DD domain due to fraction delay, and results in poor BER performance. This is evident in the BER performance shown in Fig. \ref{fig:ber}. It can be observed that at $500$ km /h the proposed TFL-OTFS outperforms RC-OTFS by $2$~dB at higher SNRs for both perfect CSI case as well as with estimated CSI. 
\begin{figure}
    \begin{subfigure}{0.5\columnwidth}
    \includegraphics[width=1\columnwidth]{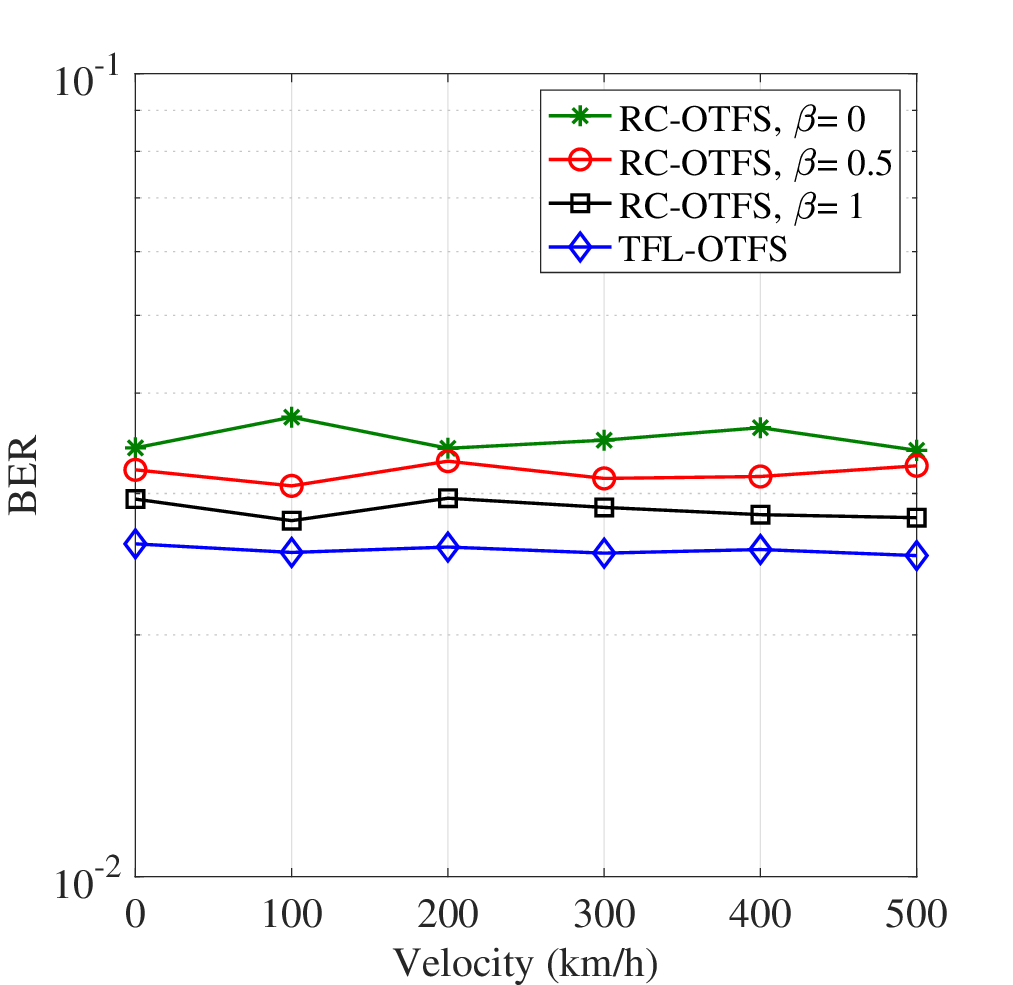}
    \caption{NMSE}
    \label{fig:nmse_var}
\end{subfigure}
\hspace{-0.3 cm}
   \begin{subfigure}{0.5\columnwidth}
    \includegraphics[width=1\columnwidth]{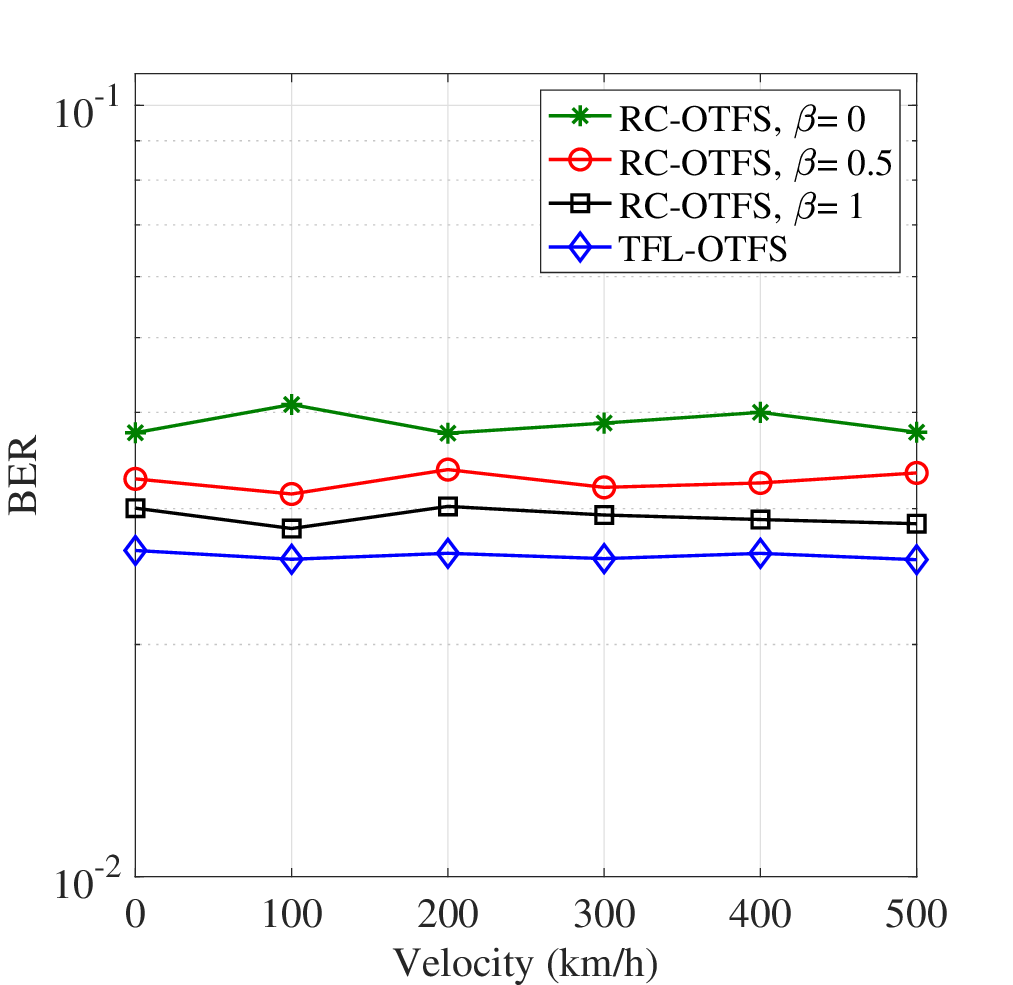}
    \caption{BER}
    \label{fig:ber_var}
\end{subfigure}
    \caption{Channel estimation NMSE and BER performance versus relative velocity of the transmitter and receiver.}
    \label{fig:var}
\end{figure}

To evaluate the performance of the proposed TFL pulse for different values of Doppler spread, we show the BER performance of the MMSE detector at $15$ dB SNR as a function of mobile speeds in Fig. \ref{fig:ber_var}. It can be observed that the TFL-OTFS always gives better performance in terms of NMSE and BER for mobile velocities in the range of $0$ to $500$ km/h. It is to be noted that both RC-OTFS and TFL-OTFS are equally non-sensitive to increasing Doppler spread. 

\begin{figure}
    \begin{subfigure}{0.5\columnwidth}
    \includegraphics[width=1\columnwidth]{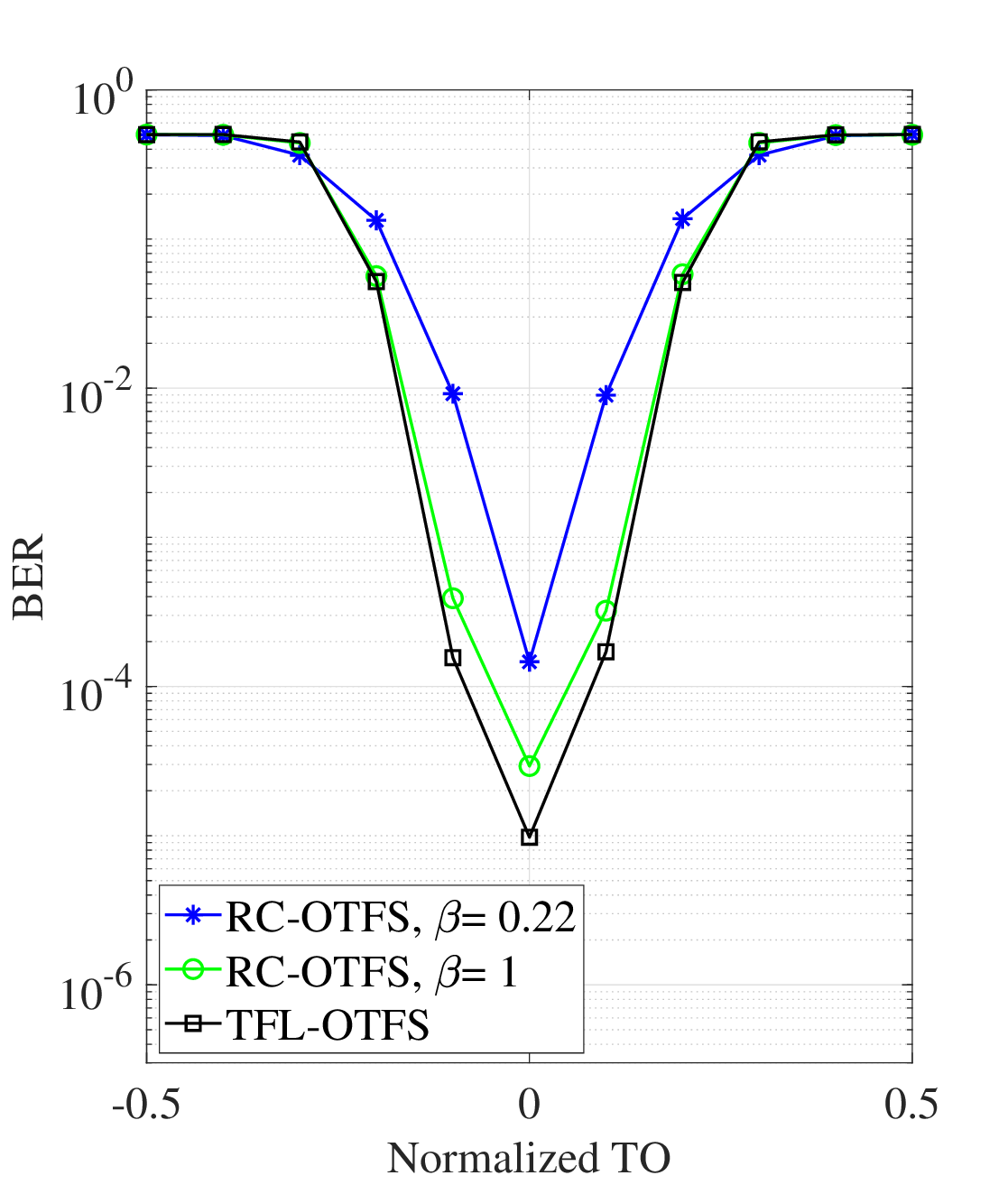}
    \caption{AWGN channel}
    \label{fig:to1}
\end{subfigure}
\hspace{-0.3 cm}
   \begin{subfigure}{0.5\columnwidth}
    \includegraphics[width=1\columnwidth]{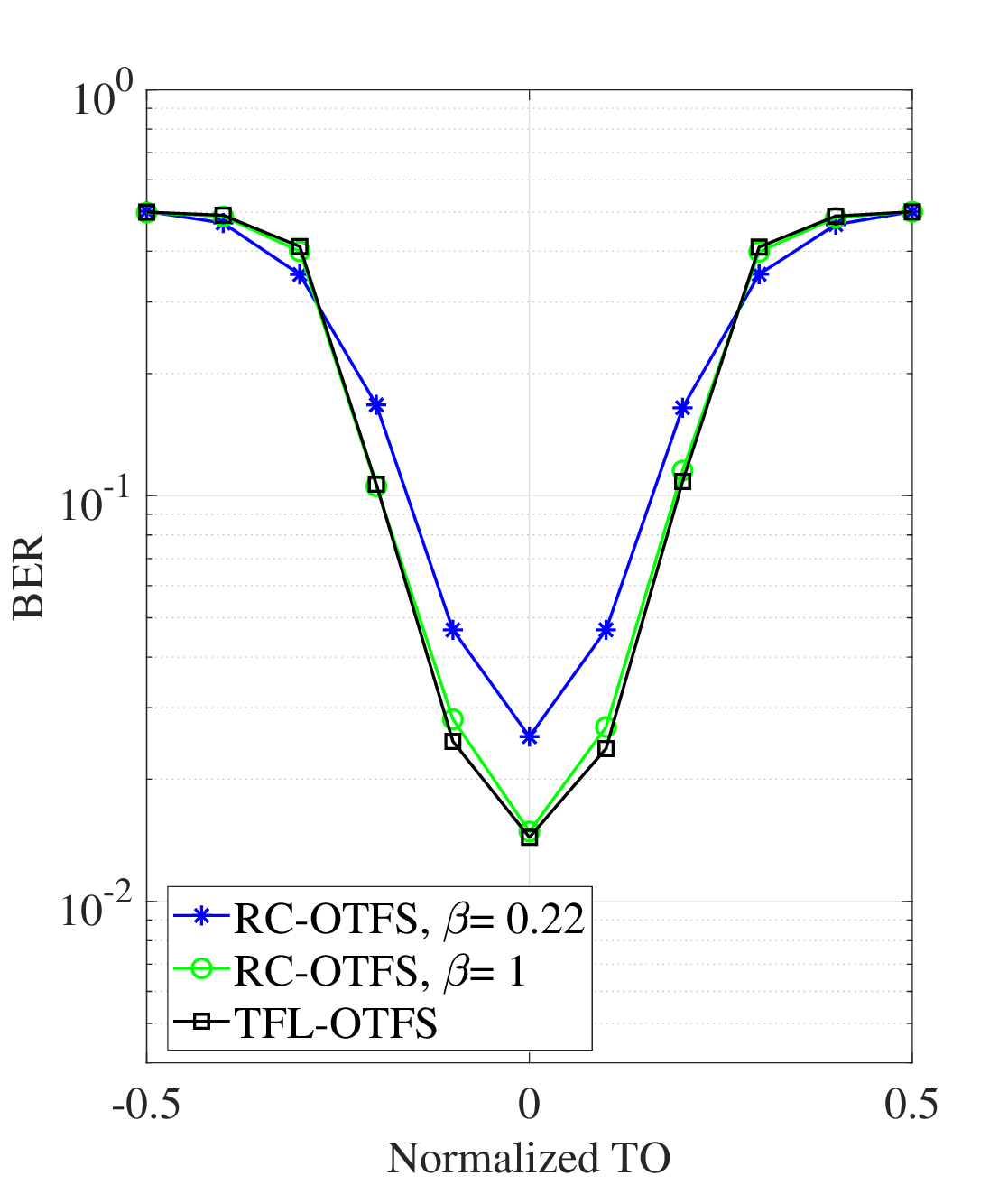}
    \caption{LTV channel}
    \label{fig:to2}
\end{subfigure}
    \caption{Comparison of TFL- and RC-OTFS for different TOs.}
    \label{fig:to}
\end{figure}

In addition to the above simulation setting, we also investigated the robustness of the proposed TFL-OTFS to uncompensated TOs. Even though the integer part of the TO normalized to $T_{\rm s}$ can be estimated, the fractional part can result in residual TO. The uncompensated fractional TO results in performance degradation. 
Fig. \ref{fig:to1} shows the BER performance of TFL-OTFS and RC-OTFS for different values of fraction TOs in an AWGN channel with $20$ dB SNR. It can be seen that RC-OTFS is more sensitive to uncompensated fractional TO than the TFL-OTFS. We also analyzed the robustness to TO in linear time-varying (LTV) channel with mobile velocity of $500$ km/h with no fractional delays and the BER results are shown in Fig. \ref{fig:to2}. It can be observed that even in an LTV channel the proposed TFL-OTFS outperforms the RC-OTFS. 
\section{Conclusion}\label{sec:conl}
In this paper, we developed a vector matrix input-output relationship for DD domain data transmission and analyzed the interference due to fractional delay and Doppler. For the first time in the literature, we proposed the use of a TFL pulse for DD domain data transmission. We showed that the traditional RC-OTFS fails to exploit the full advantages offered by the DD domain due to the presence of fractional delays. Simulation results demonstrate that the proposed TFL-OTFS outperforms the RC-OTFS in the presence of fractional delays. Furthermore, the TFL-OTFS is more robust to uncompensated fractional TOs compared to RC-OTFS. The localization of the TFL-OTFS across the delay dimension makes it a suitable candidate for joint communication and sensing applications in next-generation wireless communication systems. 
\bibliographystyle{IEEEtran}
\bibliography{Main_TF_Hermite_ref}
\end{document}